\documentclass[12pt,preprint]{aastex}

\newcommand{\ltsim}{\lower.5ex\hbox{$\; \buildrel < \over \sim \;$}}
\newcommand{\gtsim}{\lower.5ex\hbox{$\; \buildrel > \over \sim \;$}}

\newcommand{\gapprox}{\lower.4ex\hbox{$\;\buildrel >\over{\scriptstyle\sim}\;$}}
\newcommand{\lapprox}{\lower.4ex\hbox{$\;\buildrel <\over{\scriptstyle\sim}\;$}}

\def\fluxunit   	{ergs cm$^{-2}$s$^{-1}$}

\def\pcufluxunit 	{counts s$^{-1}$PCU$^{-1}$}

\newcommand{\Msun}      {\mbox{$\,{\rm M}_{\mathord\odot}$}}

\newcommand{\Mdot}{\mbox{$\dot M$}}

\newcommand{\exo}{\mbox{EXO\,0748$-$676}}

\slugcomment{Submitted to Astrophysical Journal}

\begin{document}

\title{A Strong X-Ray Burst from the Low 
Mass X-Ray Binary EXO~0748--676}

\author{Michael T. Wolff\altaffilmark{1},}
\affil{Space Science Division, 
Naval Research Laboratory, 
Washington, DC 20375}

\author{Peter A. Becker\altaffilmark{2},}
\affil{Center for Earth Observing and Space Research,
George Mason University,
Fairfax, VA 22030-4444}

\author{Paul S. Ray\altaffilmark{3}, 
Kent S. Wood\altaffilmark{4}}
\affil{Space Science Division, 
Naval Research Laboratory, 
Washington, DC 20375}

\altaffiltext{1}{E-mail Address: Michael.Wolff@nrl.navy.mil}
\altaffiltext{2}{E-mail Address: pbecker@nrl.navy.mil}
\altaffiltext{3}{E-mail Address: Paul.Ray@nrl.navy.mil}
\altaffiltext{4}{E-mail Address: Kent.Wood@nrl.navy.mil}

\begin{abstract}
We have observed an unusually strong X-ray burst 
as a part of our regular eclipse timing 
observations of the low mass binary system \exo.
The burst peak flux was $5.2 \times 10^{-8}$ 
\fluxunit, approximately five times the normal 
peak X-ray burst flux observed from this source 
by RXTE.
Spectral fits to the data strongly suggest that 
photospheric radius expansion occurred during 
the burst.
In this Letter we examine the properties of this 
X-ray burst, which is the first example of a radius 
expansion burst from \exo\ observed by RXTE.
We find no evidence for coherent burst oscillations.
Assuming that the peak burst luminosity is the 
Eddington luminosity for a 1.4 \Msun\ neutron star 
we derive a distance to \exo\ of $7.7$ kpc for a 
helium-dominated burst photosphere and $5.9$ kpc 
for a hydrogen-dominated burst photosphere.
\end{abstract}


\section{Introduction}

The low mass X-ray binary (LMXB) system \exo\ was 
originally discovered by EXOSAT during February 1985
\citep{pwgh85}.  
Observations by EXOSAT detected periodic full 
eclipses and 25 Type I X-ray bursts, three of 
which were reported to show evidence of photospheric 
radius expansion~\citep{ghpw86}.  
\exo\ was subsequently observed by GINGA, ROSAT, ASCA, 
Beppo-SAX, XMM-Newton, and the Rossi X-Ray Timing 
Explorer (RXTE). 
\exo\ is a 3.82 hr orbital period system, normally 
classified as an ``atoll'' source, with a neutron star 
primary accreting matter from the Roche lobe-filling, 
low-mass main-sequence secondary star UY Vol.  
A possible observation of gravitationally red-shifted 
lines in X-ray burst spectra has been reported 
by \citet*{cpm02}.  
\citet{vs04a} have recently reported a possible spin 
period for the neutron star of 45 Hz.

\exo\ is one of the most heavily studied 
and important LMXB physics laboratories so constraining 
the source distance as accurately and rigorously 
as possible is very important.
The spectral type of the secondary in \exo\ has, so far, 
been elusive. 
An upper limit for the U$-$brightness of the secondary star 
of 21$^{\rm st}$ magnitude during X-ray quiescence can 
be estimated based on the absence of 
detectable emission on the SRC-J survey plate 
of the \exo\ region \citep{wqhm85}.
Observations of \exo\ after its discovery by EXOSAT 
during its X-ray active state show an optically 
erratic variable of near 17$^{\rm th}$ 
magnitude \citep*{pkp88}.
However, \citeauthor{pkp88} attributed the bulk of the 
optical emission to the neutron star accretion disk.
The intrinsic faintness of the secondary and the system 
location outside the galactic bulge 
at $({\rm l},{\rm b}) = (280.0^{\circ},-19.8^{\circ})$, 
makes an accurate reddening distance difficult to 
obtain~\citep{sz90}. 
Furthermore, its galactic latitude puts it well out of 
the galactic plane making a galactic differential 
rotation distance uncertain by an unknown 
amount~\citep{ccs+86}.
It is also too distant for the 
present generation of astrometry satellites to measure 
an accurate parallax.
Yet the fact that \exo\ is a bursting LMXB opens the 
possibility of obtaining the distance if an Eddington 
luminosity-limited X-ray burst can be observed
and accurately analyzed. 
\citet{ghpw86} analyzed EXOSAT observations of three 
apparent radius expansion bursts but only assumed a 
distance of 10 kpc to \exo\ in order to estimate the 
Eddington luminosity.
\citet{jn04} used the \citeauthor{ghpw86}
peak burst flux from the EXOSAT Medium Energy detector 
in order to estimate the distance of \exo\ to be 
in the range 6.8--9.1$\,$kpc.
Until now, however, no X-ray burst detected by RXTE 
from \exo\ showing evidence of photospheric radius 
expansion has been reported. 

Since 1996, near the beginning of the RXTE mission, we 
have maintained a program of regular observations 
of \exo, of length 2--3$\,$ks, in order to monitor 
the eclipses and precisely time the system orbital 
period \citep{hwc97,whw+02}.
During our eclipse monitoring observations with the 
Proportional Counter Array 
(PCA) on RXTE we serendipitously capture numerous 
Type I X-ray bursts. 
A typical X-ray burst will rise within seconds to an 
apparent count rate of roughly 
1000 \pcufluxunit\ in the 2--20$\,$keV energy 
range ($\sim 1/3$ Crab).
However, during May 2004 we observed one X-ray burst 
that reached a count rate of nearly 
5000 \pcufluxunit\ ($\sim 1.7$ Crab)
and showed evidence of photospheric radius expansion.
In this paper we report a detailed analysis of this 
very luminous burst. 
Assuming the peak burst luminosity is equal to the 
Eddington luminosity for a 1.4 \Msun\ neutron star, 
we derive a distance to the \exo\ system of $7.7\,$kpc 
for a helium-dominated burst photosphere, and $5.9\,$kpc
for a hydrogen-dominated burst photosphere.

\section{RXTE Observations}

During RXTE observation designated by the ObsID 
90059-03-01-00, we observed a strong X-ray burst 
that began at MJD (TT; not barycentered) = 53126.40358.
Four PCUs (PCU0, PCU1, PCU2, PCU3) were active 
during this observation.
The observation was done in {\it GoodXenon} mode 
which preserves full time and spectral information 
for each photon event. 
The burst peak flux saturated the {\it GoodXenon} 
data collection buffer twice during the burst, 
triggering the burstcatcher mode, which preserves 
only the event times binned at 122 $\mu$s intervals and 
no spectral (PHA channel) or layer information.
Time-dependent spectral analysis of the burst can 
be performed only in sections of the observation where 
there are no gaps in the {\it GoodXenon} data.
However, we can construct a continuous light curve 
for the entire X-ray burst using {\it Standard-1} data, 
which has no gaps.

For the spectral analysis of this burst layer 1 
events were extracted from the PCA FITS files for 
which unsaturated {\it GoodXenon} event data is
available.  
We utilize version 5.3.1 of the HEASOFT FTOOLS package
with XSPEC version 11.3.1 for the spectral fitting.  
This version of the FTOOLS generates response matrices 
with updated values of the default geometric areas 
for each PCU \citep{jmr+05}.
These new effective areas improve the flux calibration 
of the Crab pulsar plus nebula so that the Crab flux 
is consistent with the standard value.  
This makes adjustments to the flux such as those utilized 
by \citet{kdi+03} no longer necessary.  
Deadtime corrections are applied (always less 
than 7.6\%) to the observed count rates and spectra.

\begin{figure}
\centerline{\includegraphics[height=6.0in,angle=270.0]{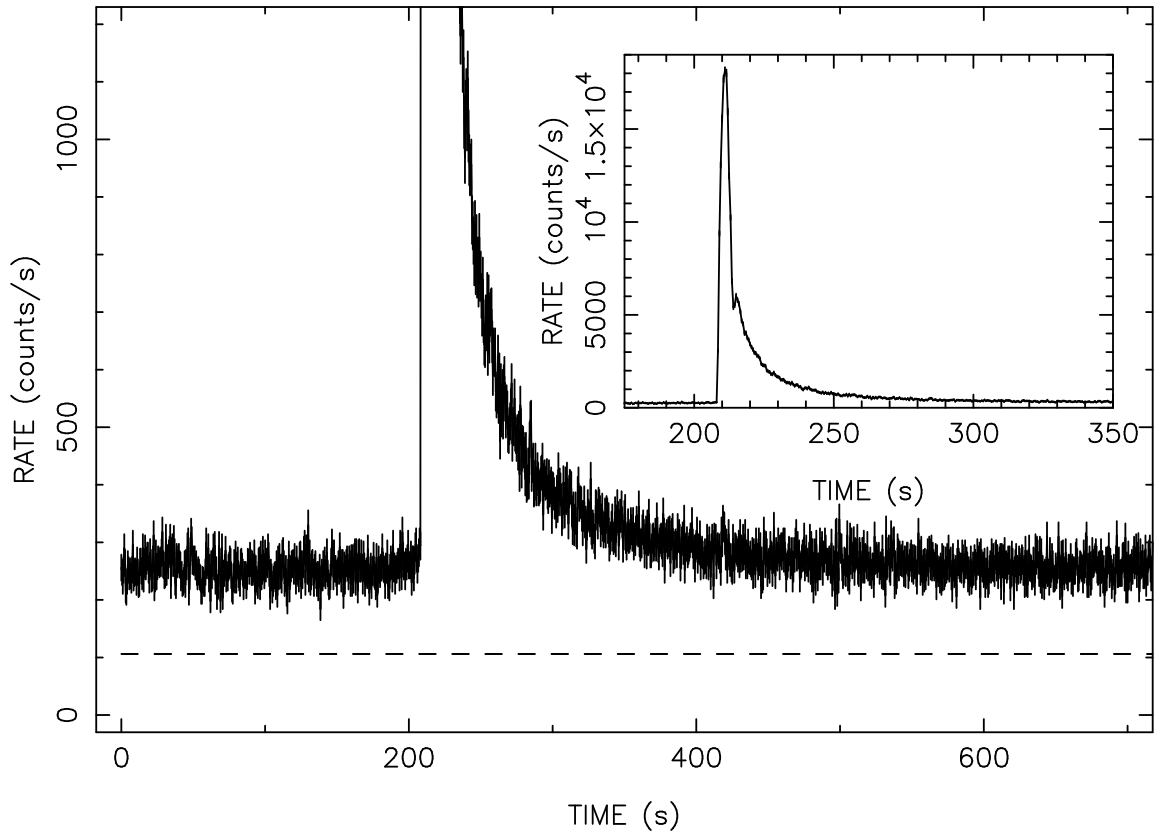}}
\caption{
The {\it Standard-1} light curve of the X-ray burst from the
four active PCUs.
The dashed line shows the approximate background count level
in the PCA during the burst.
The pre-burst light curve shows only a small level of variation
consistent with little or no dipping behavior as noted in 
the text.
\label{fig-standard1}}
\end{figure}

We fix the persistent X-ray emission for the source by 
extracting the spectrum of the 200 seconds of data 
before the X-ray burst began and use this as a fixed 
background for the burst spectral analysis.
The {\it Standard-1} mode light curve is shown in
Figure~\ref{fig-standard1}. 
Examination of Figure~\ref{fig-standard1} shows 
that this pre-burst segment of data has relatively 
little flickering or dipping behavior that might  
signify locally intervening 
material in the \exo\ system and we therefore ignore 
this effect in our analysis.
We fit the {\it GoodXenon} spectral data derived from 
segments of the burst profile to a model consisting 
of the XSPEC blackbody emission function multiplied 
by a photoelectric absorption function: 
{\it wabs$\times$bbodyrad}.
The interstellar hydrogen column density is frozen 
at $4.0 \times 10^{21}$ cm$^{-2}$ obtained 
by \citet{tcs+97} from ASCA measurements. 
This allows the fitted parameters for the X-ray 
burst itself to vary independently of 
the persistent emission.
For our spectral fits we include counts in the 
energy range 2.0$-$20 keV.
We generate one response matrix for the entire X-ray 
burst observation which corresponds to approximately 100 
seconds of time, and we
assume zero systematic error in the fitting procedure.
The resulting reduced $\chi^2$ values for the spectral 
fits are all in the range $0.5-2.0$.
This procedure is similar to the analysis method 
of \citet{smal01} for an X-ray burst observed from 
X2127+119 in the globular cluster M15.
As a test of our method, we analyzed the RXTE archival 
data for the same X-ray burst observed by \citet{smal01} 
and found results similar to his, which confirms that
our analysis procedure is reliable.

\section{Burst Oscillation Search}

The unusual strength of this burst makes it particularly 
attractive for burst oscillation searches.  
\citet{vs04a} found evidence for a 45-Hz pulsation 
frequency in an incoherent sum of power spectra from 
38 bursts from \exo\, but oscillations have never been 
detected in a single burst.  
We searched both the {\it GoodXenon} and burstcatcher 
data for coherent oscillations during this strong burst.

The burstcatcher data were binned at 122 $\mu$s intervals
for 10 s starting about 1 s before the beginning of 
the burst.  
These data are continuous, but have no energy resolution.  
The {\it GoodXenon} data have some gaps during the 
brightest part of the burst because the telemetry buffers 
filled.
However, full energy resolution was available and 
we searched 100 seconds of data beginning at 0.5 s 
before the burst.   
We created three time series to search for pulsations, 
one from the burstcatcher data with 122 $\mu$s bins and 
two from the {\it GoodXenon} data with 244 $\mu$s bins 
with energy selections of 2.5--6.0$\,$keV and 6.0--60.0$\,$keV.
All three time series were searched for oscillations using 
the \texttt{powspec} FTOOL.
A third-order polynomial was removed from each data 
window to reduce the red noise from the overall burst 
profile, and power spectra were Leahy normalized.  
Several combinations of FFT window length (from 2048 to 
32768 points) and numbers of power spectra averaged 
(from 1 to 10) were searched to maximize 
sensitivity to oscillations that occur with various 
timescales.

No oscillations were found during the X-ray burst.  
The precise upper limit on the oscillation amplitude 
depends on the assumed duration of the oscillations, 
but for averages of four 8192-point (2 second) intervals 
in the 6.0--60$\,$keV time series, we would easily have 
detected sinusoidal pulsations whose RMS was at 
least 5\% of the total RMS fluctuations.
However, we would not have detected burst oscillations 
with an amplitude of only 3\% such as those found by 
\citet{vs04a}.

\section{Burst Spectral Analysis}

In our discussion of the burst physics we adopt the 
notation of \citet*{ehs84}, and define the observed 
``blackbody radius'' $R_{\rm bb}$ with respect 
to the bolometric luminosity $L$
and the observed color temperature $T_{\rm c}$ using
\begin{equation}
L = 4 \pi R_{\rm bb}^2 \, \sigma T_{\rm c}^4
\ .
\label{eq0.1}
\end{equation}
Solving for the blackbody radius yields
\begin{equation}
R_{\rm bb} = \left(L \over 4 \pi \, \sigma T_{\rm c}^4\right)^{1/2}
\ .
\label{eq0.2}
\end{equation}
In the \citet{ehs84} development, the color temperature 
observed at infinity $T_{\rm c}$ is the physical temperature 
of the radiating plasma at the ``color radius'' $r_{\rm c}$, 
where thermalization of the radiation occurs via either 
free-free absorption or Compton scattering.
In this sense, the color radius can be interpreted as 
the standard photospheric radius. 
On the other hand, $R_{\rm bb}$ is not a true physical 
radius but rather a parameter that ensures that the 
observed spectral flux is correctly fitted in 
the {\it bbodyrad} model. 
\citeauthor{ehs84} showed that $R_{\rm bb}$ is related 
to the color radius via $r_{\rm c} = S\,R_{\rm bb}$, 
where $S = (3 \tau_{\rm c}/4)^{1/2}$ is 
the ``radius expansion factor'' and $\tau_{\rm c}$ is 
the electron scattering optical depth to the color 
radius \citep*[see also][]{lth86,smal01}. 
Physically, the distinction between the two radii arises 
as a consequence of electron scattering. 
The presence of a relatively deep scattering layer 
above the color radius reduces the diffusion velocity 
of the photons, which in turn reduces the outgoing flux. 
In order to carry the required luminosity, $r_{\rm c}$ must
therefore exceed
the value of $R_{\rm bb}$ computed using 
equation~(\ref{eq0.2}), since this expression completely
ignores the effect of diffusion by implicitly assuming 
that the radiation is directly emitted from 
the ``naked'' stellar surface. 
Hence $R_{\rm bb}$ is significantly smaller than the 
true physical radius $r_{\rm c}$ during the 
scattering-dominated, brightest part of the burst. 
This point is further discussed below.

The value of the Eddington luminosity $L_\mathrm{Edd}$ 
observed at infinity is given by the standard expression as 
modified by~\citet*{lvt93} and~\citet{gpcm03},
\begin{equation}
L_{\rm Edd} = {4 \pi GM_{*}c \over \kappa_{\rm sc}}
\left( 1 - {{2GM_{*}} \over {c^2r_{\rm c}}} \right)^{1/2}
\left[ 1 + (2.2 \times 10^{-9} T_{\rm c})^{0.86} \right] \, ,
\label{eq-LEdd}
\end{equation}
where $M_{*}$ is the neutron star mass (taken to be 
1.4 \Msun) and $\kappa_{\rm sc}$ is the electron 
scattering cross section.
The temperature-dependent factor in brackets in 
equation~(\ref{eq-LEdd}) expresses the high temperature 
modification of the scattering cross section.
This set of general relativistically-correct 
relations allows us to determine the distance 
to \exo.

\begin{figure}
\centerline{\includegraphics[height=7.0in,angle=0.0]{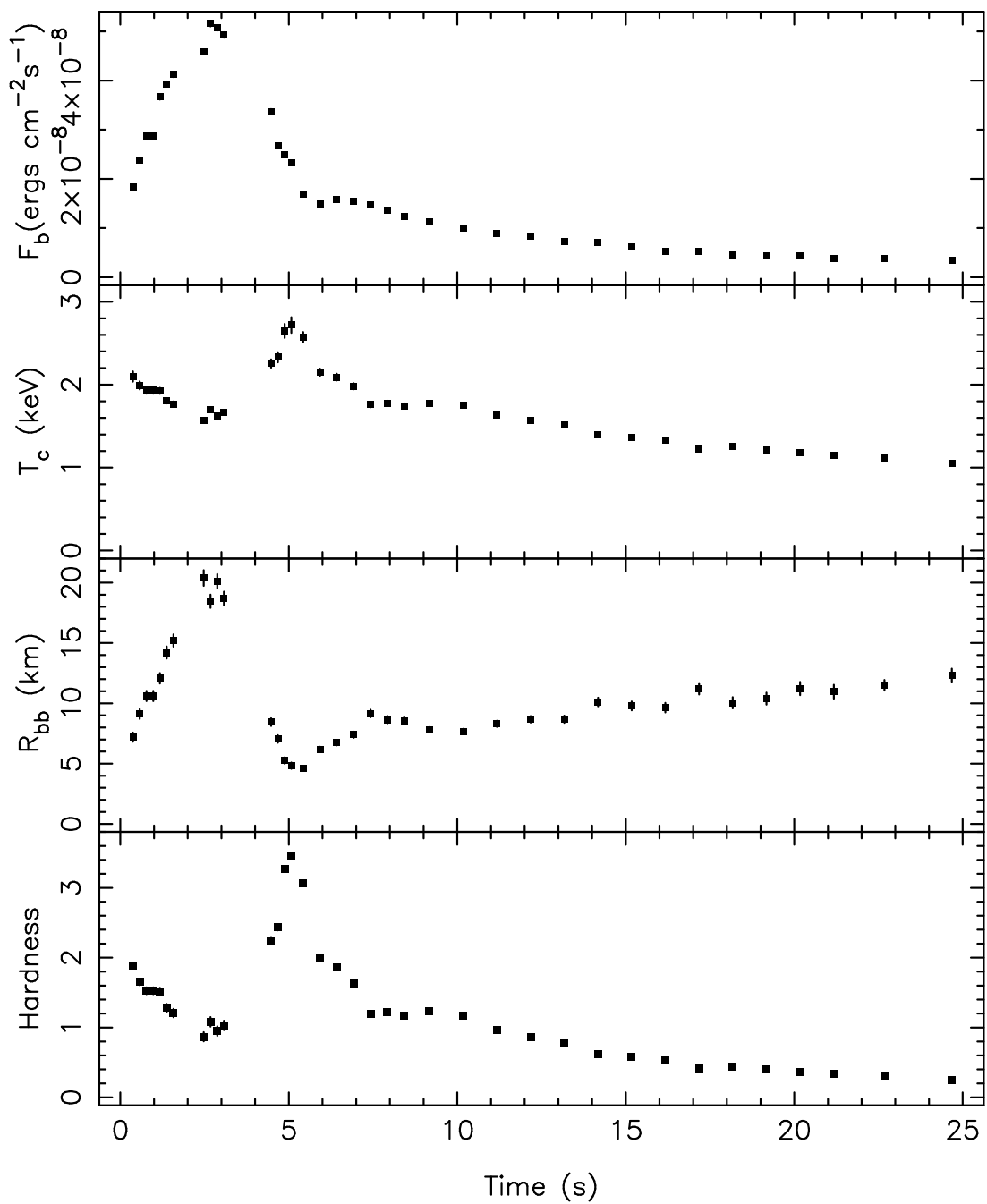}}
\caption{
Results for the bolometric flux, observed color temperature, 
and blackbody radius from the spectral fit during the 
X-ray burst assuming a distance of 7.7 kpc.
Note the gaps in the data points beginning at 1.6 s and 3.1 s.
These correspond to the saturation of the {\it GoodXenon} 
event mode.
The peak bolometric flux 
is $F_\mathrm{peak} = 5.16 \times 10^{-8}$ ergs cm$^{-1}$ s$^{-1}$.
\label{fig-fourpanel}}
\end{figure}

The X-ray burst takes $\sim 2.5$ s to rise to its peak count
rate in the light curve shown in Figure~\ref{fig-fourpanel}. 
This rise takes the observed flux from a persistent 
source-plus-background count rate of 276 counts s$^{-1}$ 
in the 2--20$\,$keV energy range ($\sim 16$ mCrab) to a 
burst peak rate over 18,000 counts s$^{-1}$.
The \exo\ orbital phase at the start of the burst is 
$\phi = 0.034220(4)$, roughly 230 s after eclipse egress.
After this the flux spends $\sim 1.7$ s within 90\% of 
the peak count rate.
The flux then falls rapidly for $\sim 2$ s down to a plateau 
at a level $\sim 30\%$ of the peak rate and then starts
a rapid exponential decay with an $e$-folding time of 14 s. 
The principal parameters we derive for this burst are 
given in Table~\ref{tbl-burst}.

In Figure~\ref{fig-fourpanel} the results of our
spectral fitting are presented for the helium-rich 
photosphere case.   
The hardness ratio we plot is the ratio of the flux in 
the 6--20$\,$keV energy band to the flux in the 3--6$\,$keV 
energy band.
When the observed flux reaches its peak, the
apparent blackbody radius $R_{\rm bb}$ also reaches its
maximum value as seen in Figure~\ref{fig-fourpanel}. 
As the observed blackbody radius initially expands 
to $\sim 20$ km, the observed color temperature 
decreases to a local minimum at the moment 
the radius achieves its largest value. 
In a similar fashion, the hardness ratio falls reaching 
a local minimum as the observed flux is reaching its peak. 
The hardness ratio then dramatically increases again 
reaching its maximum value $\sim$2 seconds after the 
flux has begun to decay.
The decrease in both the temperature and the hardness
ratio while the flux is near its peak is indicative 
of the cooling that occurs during the
photospheric expansion~\citep{sb03}.
After the flux reaches its peak and begins to decay, the 
radius drops rapidly to an apparent minimum of $\sim$4.6 km
and then recovers and asymptotically approaches $\sim$13 km.
As $R_{\rm bb}$ decreases after the peak, the temperature 
quickly rises to its largest values just as the flux is 
also decreasing.
When the radius reaches its minimum value, just after the 
peak of the burst, the temperature and the hardness ratio 
each reach their respective maxima.
After this the burst begins its exponential decay 
until it fades away completely.

We derive an unabsorbed burst fluence 
of $3.60 \times 10^{-7}$ ergs cm$^{-2}$ by multiplying 
the unabsorbed flux in the energy range 2.0--20.0$\,$keV 
by a bolometric correction and then integrating 
over the duration of the burst rise and decay.
The bolometric correction is computed as a function 
of time by dividing the total blackbody flux by the 
blackbody flux in the 2.0--20.0$\,$keV energy range based 
on the sequence of observed color temperatures.
Assuming that the expanding envelope is helium-dominated, 
and that the peak flux of $5.16 \times 10^{-8}$ ergs 
cm$^{-2}$ s$^{-1}$ corresponds to the Eddington luminosity 
of $3.64 \times 10^{38}$ ergs s$^{-1}$ for a 1.4 \Msun\ 
neutron star, we derive a distance of 7.67 kpc to \exo.
On the other hand, if the burst photosphere has solar 
abundances, then the peak flux corresponds to an Eddington 
luminosity of $2.12 \times 10^{38}$ ergs s$^{-1}$ for a 
1.4 \Msun\ neutron star, and a distance of 5.85 kpc to \exo.
The associated values of $L_\mathrm{Edd}$ for hydrogen- 
and helium-rich photospheres are consistent with the 
peak luminosities obtained by \citet{kdi+03} in their 
analysis of photospheric radius expansion bursts in 
globular cluster sources, which have known distances.
We note that the peak flux we obtain is $\sim 26$\% higher 
than the apparent peak flux obtained by \citet{ghpw86} 
for three photospheric radius expansion bursts 
from \exo\ observed by EXOSAT. 
\citet{gpcm03} found that in photospheric radius 
expansion bursts from the LMXB 4U~1728$-$34 the peak 
fluxes were not constant but instead exhibited 
variations with a RMS deviation of 9.4\% and a 
total variation of 46\%.
\citeauthor{gpcm03} concluded that this variation might 
be caused by variable obscuration from a warped 
accretion disk around the neutron star.
If the accretion disk surrounding the neutron star 
in \exo\ is also warped, then this could cause the 
same type of variations in peak fluxes as observed in
4U~1728$-$34, which could explain why our peak flux
is higher than that of \citet{ghpw86}.
If our higher flux is a result of less obscuration by 
a warped accretion disk then the distance we derive 
will be more accurate than a distance based on 
the lower EXOSAT flux because our peak flux is more 
indicative of the unobscured Eddington flux for \exo.

The \exo\ system is located at galactic latitude 
$-19.8^{\circ}$ and galactic longitude $280.0^{\circ}$. 
Hence if we assume that the galactic center is 8.5 kpc 
from the solar system, then \exo\ is $\sim10.5\,$kpc 
from the galactic center and $\sim 2.6\,$kpc below 
the galactic plane.
The ratio of the persistent X-ray flux from the
pre-burst spectrum
to the Eddington flux at the peak of the burst is 
$\gamma=0.0094$ (see Table~\ref{tbl-burst}),
which is consistent with the assumption that \exo\ is
an atoll LMXB source. 
The distance implied by the helium-rich 
photosphere case combined with the observed persistent 
X-ray flux yields a mass accretion rate 
of $\sim 2.9 \times 10^{-10}\,$\Msun yr$^{-1}$, which is 
in the accretion rate range expected to yield mixed hydrogen 
and helium burning triggered by thermally unstable
helium ignition~\citep{sb03}.
On the other hand, if the burst photosphere is hydrogen-rich
then the closer distance implies a lower persistent \Mdot,
pushing \Mdot\ into the range that results in burst ignition via
thermally unstable hydrogen burning~\citep{sb03}.
Thus, conclusions about the characteristics of the burning 
for this X-ray burst depend on the assumed composition  
of the burst photosphere.

A number of possible sources of error exist for our 
estimated distance to \exo.
First, if the assumed mass of the neutron 
star ($M_{*} = 1.4\,$\Msun) is 
incorrect then the Eddington luminosity has been incorrectly
calculated and the distance must be adjusted accordingly.
For a 1.6 (1.2) \Msun\ neutron star we derive a distance of 
8.2 (7.1) kpc from the same spectral fitting data 
for a helium-dominated burst.
A second source of error enters through our estimate of
the burst bolometric flux, either due to errors in
the PCA calibration or errors in the bolometric corrections
derived from the fitted temperatures.
If the flux is 20\% too high (low) we estimate that
the source distance would change to 7.0 (8.6) kpc,
again for a helium-dominated burst. 
The error estimates for the hydrogen-rich burst case are
similar.

The increase in $R_{\rm bb}$ during the early phase of the 
burst clearly suggests that the photospheric radius of the 
radiating atmosphere expands. 
However, the subsequent minimum value 
of $R_{\rm bb} \sim 4.6\,\,$km just after the burst peak 
is only about one third of the apparent radius that 
a distant observer would associate with a 1.4 \Msun\ neutron 
star with a proper radius of 10 km, which is about 13 km 
when the gravitational redshift is included \citep{st83,lp01}. 
If one expects a proper neutron star radius of $\sim$ 10 km, 
then that in turn suggests that the {\it bbodyrad} model 
in XSPEC does not include all of the physics involved in 
producing a radius expansion X-ray burst light curve. 
Without a clear understanding of the physics involved 
in the light curve production after the peak of the burst, 
the derived radii cannot be reliably 
used to constrain either the neutron star radius or 
the equation of state of the nuclear matter 
\citep[see the discussion of this point in][]{lvt93}.
During the early, high-temperature portion of the burst
around the time of peak flux, 
the thermalization of the emitted radiation is almost 
certainly dominated by Comptonization. 
In this situation, the scattering depth to the color 
radius, $\tau_{\rm c}$, is related to the observed color 
temperature, $T_{\rm c}$, 
via $\tau_{\rm c} = (m_e c^2/4kT_{\rm c})^{1/2}$ 
\citep{lth84,lth86,smal01}. 
Based on the observed color temperature at the peak of 
the burst, we find that $\tau_{\rm c} \sim 5$, and therefore 
$S \sim 2$, where $S$ is the radius expansion factor defined 
above. 
This implies that the color radius $r_{\rm c}$ is roughly 
double the blackbody radius just after the burst peak, 
when $R_{\rm bb}$ achieves its minimum value. 
However, even a factor of two increase in $R_{\rm bb}$ is 
not sufficient to bring the color radius $r_{\rm c}$ into 
accord with the apparent radius of 13 km.
It should be noted that the canonical value of 10 km for the proper 
radius of the neutron star is based on a particular equation of 
state for the nuclear matter, and some alternative equations 
may allow slightly more compact neutron stars \citep{hp00}, with 
proper radii perhaps as small as 9 km.
However, this adjustment does not appear to be sufficient
to completely remove the discrepancy cited above.

With regard to the post-peak-flux drop in $R_{\rm bb}$ it is
possible that non-thermal spectral distortions could be 
created by ionized material located along the line of sight. 
Such material could rise from the accretion disk due to 
strong heating and photoionization driven by the peak 
X-ray flux. 
Preferential absorption of the soft radiation by the 
ionized gas would decrease the total flux, while also causing 
the apparent color temperature to rise, in agreement with
the observations (see Figure~\ref{fig-fourpanel}). 
These two effects will cause a reduction in the value 
of $R_{\rm bb}$ computed using equation~(\ref{eq0.2}). 
In this sense, the X-ray burst will have created its 
own ``weather'' that it must shine through if the photons 
are to reach the PCA. 
There is some evidence for the existence of ionized gas 
in the vicinity of the binary system 4U 1820$-$303 during 
X-ray bursts although no detailed model for this phenomenon 
has appeared \citep{sb02}. 
Presumably, after the early phase of strong irradiation, 
the ionized material drops back down into the disk and 
therefore the differential absorption effect disappears.
However, it is not clear if the dynamical and thermal 
timescales in the gas support this interpretation.

\section{Conclusions}

We have analyzed an unusually strong X-ray burst from the 
LMXB \exo\ and found that it shows evidence for 
photospheric radius expansion. 
No oscillations were detected during the X-ray burst
despite an exhaustive search. 
When we equate the peak flux from the X-ray burst 
with the expected Eddington flux from a 1.4 \Msun\ 
neutron star we derive a distance to 
the \exo\ system of $7.7$ kpc for a helium-dominated 
burst photosphere and $5.9$ kpc for a hydrogen-dominated 
burst photosphere.
While we have argued that our distance determination 
to \exo\ is likely to be more accurate than previous 
determinations the ambiguity in the distance determinations
between hydrogen- and helium-rich photospheric abundances 
in the real Eddington luminosity remains. 
Thus, we must quote a range of distances accounting
for both abundance cases. 
Only when the observational characterizations of burst 
properties are able to fix the abundances of the bursting 
neutron star photosphere can this ambiguity in distance 
determinations to bursting LMXB systems be overcome. 
The X-ray burst flux light curve strongly suggests 
that we do not completely understand the physics of 
the spectral formation process in an X-ray burst just 
after maximum brightness, when the luminosity is 
beginning to recede from the Eddington limit. 
If the derived distance is correct, then the small values
obtained for the fitted blackbody 
radius $R_{\rm bb}$ after the peak of the 
burst are difficult to reconcile with the 10 km 
proper radius of the star, even
when adjustments are made based on the high 
electron scattering optical depth. 
Possible alternative explanations include 
the presence of intervening material that 
absorbs the soft flux after the peak, or perhaps 
the occurrence of non-spherical mass motions.

\begin{acknowledgments}
We are pleased to acknowledge discussions with 
Drs. Craig Markwardt, Jean Swank, 
Deepto Chakrabarty, Duncan Galloway,
Jeroen Homan, and Lev Titarchuk.
This research is supported by the Office of 
Naval Research, the NASA Astrophysical Data 
Program, and the NASA RXTE Guest 
Observer Program.
\end{acknowledgments}

\clearpage


\clearpage

\clearpage

\begin{deluxetable}{cc}
\tablecolumns{2}
\tablecaption{\bf{\exo\ Burst and Derived Parameters}\label{tbl-burst}}
\tablehead{\colhead{Parameter} &\colhead{Value}}
\tablewidth{420pt}
\startdata
Burst Fluence ($=E_b$) &$3.60\times10^{-7}$ ergs cm$^{-2}$\\
Peak Bolometric Flux ($=F_\mathrm{peak}$) &$5.16\times10^{-8}$ ergs cm$^{-2}$ s$^{-1}$\\
Persistent Flux ($=F_\mathrm{per}$) &$4.81\times10^{-10}$ ergs cm$^{-2}$ s$^{-1}$\\
$\gamma$ $[= F_\mathrm{per}/(F_\mathrm{peak}-F_\mathrm{per})]$ &$0.0094$\\
$\alpha$ $[= F_\mathrm{per}\Delta T/E_b]$\tablenotemark{a} &$17$\\
$\tau$ $[= E_b/F_\mathrm{peak}]$ &$7.0$\\
\cutinhead{Derived Parameters Assuming Hydrogen Burst Photosphere}
\exo\ Distance From Sun\tablenotemark{b,c}&$5.9\pm0.9$ kpc\\
\exo\ Distance To Galactic Center\tablenotemark{b}&9.5 kpc\\
\exo\ Distance Below Galactic Plane&2.0 kpc\\
Burst Energy Output &$1.48\times10^{39}$ ergs\\
Persistent Accretion Rate $\Mdot_\mathrm{per}$ &$1.68\times10^{-10}$ \Msun/year \\
\cutinhead{Derived Parameters Assuming Helium Burst Photosphere}
\exo\ Distance From Sun\tablenotemark{b,c}&$7.7\pm0.9$ kpc\\
\exo\ Distance To Galactic Center\tablenotemark{b}&10.5 kpc\\
\exo\ Distance Below Galactic Plane&2.6 kpc\\
Burst Energy Output &$2.54\times10^{39}$ ergs\\
Persistent Accretion Rate $\Mdot_\mathrm{per}$ &$2.89\times10^{-10}$ \Msun/year \\
\enddata
\tablenotetext{a}{Assuming a burst recurrence time of $\Delta T = 0.285$ hours \citep{gmc+05}}
\tablenotetext{b}{Assuming a solar system distance to the galactic center of 8.5 kpc.}
\tablenotetext{c}{Assuming $L_\mathrm{peak} \,\, = \,\, L_\mathrm{Edd}$.}
\end{deluxetable}

\end{document}